# Single-exposure full-field multi-depth imaging using low-coherence holographic multiplexing

LAUREN WOLBROMSKY, NIR A. TURKO AND NATAN T. SHAKED*

*Tel Aviv University, Faculty of Engineering, Department of Biomedical Engineering, Tel Aviv 69978, Israel.*
*Corresponding author: nshaked@tau.ac.il*

**We present a new interferometric imaging approach, which allows for multiple-depth imaging in a single acquisition, using off-axis low-coherence holographic multiplexing. This technique enables sectioned imaging of multiple slices within a thick sample, in a single image acquisition. Each slice has a distinct off-axis interference fringe orientation indicative of its axial location, and the camera acquires the multiplexed hologram containing the different slices at once. We demonstrate the proposed technique for amplitude and phase imaging of optically thick samples.**

http://dx.doi.org/10.1364/OL.99.099999

The ability to image 3-D samples with optical sectioning has become of great importance and research in recent years, from studying 3-D morphology of specimens in biology and medicine [1] to investigating faults in printed circuit boards [2] and semiconductor wafers [3].

One of the most widespread methods today for 3-D imaging of optically thick samples is confocal microscopy, whereby the axial selectivity is obtained due to the use of pinholes that reject out-of-focus light. Multi-axial scanning of the sample produces thin optical sections, which are then used for 3-D reconstruction. Optical coherence tomography (OCT) does not use a physical device to reject out-of-focus light, but rather utilizes the low temporal coherence of the light source. OCT captures the interference between a reference wave and the sample wave with matching optical path length within the coherence length of the source to achieve a micron-scale sectioning ability. In full-field OCT, a camera is used for 2-D imaging, and either the reference is oscillated for acquiring multiple exposures or wavelength scanning is performed [4,5]. Since both confocal and OCT methods require scanning or multiple exposures, they are not suited for samples with dynamics faster than the scanning rate and are more prone to noise originating from vibrations and other environmental factors.

To be able to record 3-D scenes simultaneously, plenoptic imaging effectively divides the total numerical aperture to multiple micro-lenses, which view the sample from different angles [6]. The result is multiple perspectives of the sample, allowing 3-D imaging. However, the effective lateral resolution of the sample is 3-4 times worse than that of traditional microscopy. In addition, the method relies on textured features of the sample, and requires high calculation power and a relatively complex calibration process, inherently reducing throughput.

Digital holography is able to acquire the complex wavefront of the 3-D sample imaged by recording the interference from a beam interacted with the sample and a mutually coherent reference beam. The complex wavefront can then be used to perform digital propagation to various *z* planes inside the sample. In each z plane, one can still see the out-of-focus objects. Thus, the sectioning abilities of digital holography can be useful for sparse samples to some extent, dependent on iterative algorithms for rejection of out-of-focus objects [7]. Low-coherence holography was previously combined with confocal filtering for obtaining thin optical sectioning, still requiring sample scanning [8].

Off-axis digital holography allows reconstruction of the complex wavefront from a single camera exposure. Due to the off-axis holographic encoding, it is possible to multiplex additional data in co-located holograms with different fringe orientations. Alternate methods have broached the ability to multiplex different holographic images into a single multiplexed off-axis hologram. These include multiplexing different fields of view [9], multiplexing both quantitative-phase information with fluorescent images [10], as well as multiplexing wavelength channels [11], where up to six holograms can be multiplexed in a single camera exposure without camera spatial bandwidth loss [12]. Depth of field holographic multiplexing was implemented by several groups (i.e. [13,14]), but still suffering from the need for interactive algorithms for rejection of out-of-focus objects.

In this Letter, we present a new method for detecting the axial plane in holography. This is obtained by the orientation of the off-axis fringes in a multiplexed hologram. We use low coherence off-axis holography to multiplex four independent *z* planes into a single camera acquisition, without having the out-of-focus objects from different *z* planes, as typical to regular holography.

The proposed system is shown in Fig. 1. A low temporal coherence light source (LC; SuperK Extreme with SuperK SELECT by NKT) with FWHM of 6 nm centered around 633 nm defined a coherence length of 29.4 μm. This source illuminated a modified Michelson Interferometer setup, as shown in Fig. 1(a). The light is split by beam splitter BS1 into the sample and reference arms. The sample beam travels through lens L1 of focal length 200 mm, interacts with the reflective sample and is reflected back through L1, BS1, lens L3 of focal length 150 mm to the digital camera (CMOS Camera, 1280 x 1024, pixel size 5.2 μm, 8 bit, DCC1545M, Thorlabs). The reference arm continues to lens L2 of focal length 200 mm, after which the beam splits into four simultaneous reference arms by beam splitters BS2, BS3 and BS4. Each reference beam is reflected off a mirror (M1-M4) at a slightly different off-axis angle. Each reference mirror sits on a micrometer translation stage (PT1, Thorlabs) to allow for axial manipulation of the reference optical path. These reference beams are reflected back through their respective beam splitters, and directed to the camera by BS1, where an off-axis interference pattern is created. Due to the low-coherence gating dictated by the source, the slight axial displacement of each mirror results in interference from only one specific layer of the sample, which has a beam path difference that falls within the coherence length of the source. Thus, M1 interferes with layer 1, M2 with layer 2, and so on. The four interference patterns are then acquired on the camera in a single exposure.

The four-layered sample, illustrated in Fig. 1(b), was created by stacking four separate #0 cover slips (thickness: 0.085 mm to 0.115 mm), each printed with a single letter (O, M, N, I, respectively). The letters were created using lithography of 100 nm of chrome to achieve full reflectivity from the letters themselves. On the camera, we obtained a multiplexed hologram, providing spatial multiplexing of all four layers at once. Each layer creates a different off-axis interference fringe orientation, as shown in Fig. 1(c). This multiplexing is done by using the same spatial bandwidth of the camera, thus recording more information with the same number of camera pixels of a regular off-axis hologram. Since all four fringe patterns share the same camera grayscale dynamic range, the hologram SNR is expected to decrease by a factor of 2. A full model for SNR and expected sensitivity decrease is given in [15]. Note that the orientation of the fringes is determined by the relative angle of the reference beam to the sample beam. In order to achieve spatial separation in the frequency domain, whereby no layer overlaps with another, the fringe orientation from each layer must be different. Thus, the relative angle of each mirror to the sample beam is not enough, and one must ascertain that each layer angle is mutually exclusive.

In our case, the depth of field was 386 μm, and the lateral resolution was 70 μm. Since the depth of field is larger than the thickness of the sample, all reflective letters are in focus on the camera at once. Note that in this case, holographic digital focusing techniques would be rendered useless, since the entire sample is in focus. Similarly, acquiring the phase profile of the entire sample cannot help detect each layer axial distance due to phase unwrapping problems typical to such thick samples. On the other hand, we can now use the known axial locations of the mirrors in the reference arms and the coinciding orientations and frequencies of the off-axis interference fringes, as controlled by the angle of mirrors M1-M4, to detect the axial section inside the sample, thus knowing the order of the layers and their relative axial distances. It is important to note, however, that the large depth of field results in a low lateral resolution in the order of several tens of microns. The multiplexed hologram, acquired on the camera in a single exposure, can be mathematically represented at follows:

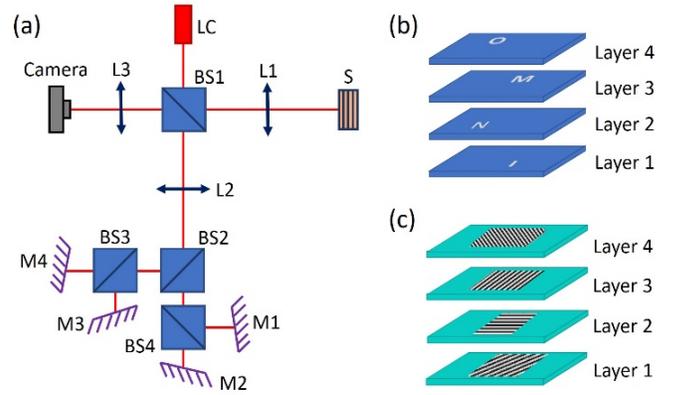

Fig. 1. (a) Schematic representation of the optical setup. LC, low-coherence light source. L1, L2, L3, lenses in 4f lens configurations. S, multilayered reflective sample. BS1-BS4, beam splitters. M1-M4, mirrors that provide the desired spatial frequencies in the off-axis interference fringes. (b) Multilayered sample made up of four coverslips, each imprinted with a reflective letter; O, M, N, I. (c) The unique fringe orientation captured from each layer of the sample from each separate reference arm, indicating the axial location of each layer.

$$I = |S_{L1} + R_1|^2 + |S_{L2} + R_2|^2 + |S_{L3} + R_3|^2 + |S_{L4} + R_4|^2$$
$$= |S_{L1}|^2 + |R_1|^2 + |S_{L2}|^2 + |R_2|^2 + |S_{L3}|^2 + |R_3|^2 + |S_{L4}|^2 + |R_4|^2 + S_{L1}R_1^* e^{j(\varphi_{S_{L1}} - \varphi_{R1})} + S_{L1}^* R_1 e^{-j(\varphi_{S_{L1}} - \varphi_{R1})} + S_{L2}R_2^* e^{j(\varphi_{S_{L2}} - \varphi_{R2})} + S_{L2}^* R_2 e^{-j(\varphi_{S_{L2}} - \varphi_{R2})} + S_{L3}R_3^* e^{j(\varphi_{S_{L3}} - \varphi_{R3})} + S_{L3}^* R_3 e^{-j(\varphi_{S_{L3}} - \varphi_{R3})} + S_{L4}R_4^* e^{j(\varphi_{L4} - \varphi_{R4})} + S_{L4}^* R_4 e^{-j(\varphi_{S_{L4}} - \varphi_{R4})},$$

(1)

where $S_{L1-4}$ are the complex wavefronts of the four layers of the sample, 1 to 4, and $\varphi_{S_{L1-4}}$ are their corresponding phases, respectively. So too, $R_{1-4}$ are the reference waves with their corresponding off-axis phases $\varphi_{R_{1-4}}$.

In the spatial-frequency domain, the first eights terms on the right side of Eq. (1) are the auto-correlation elements, representing the intensities of the recorded waves, and located around the center of the spatial-frequency domain. The remaining eight terms represent the cross-correlation elements between the waves. Note that due to the coherence gating of the system, the displacement of the mirror in each reference wave ensures that there are no cross interferences between the sample layers or between the reference waves, as each mirror in the reference arm is located at a distance greater than the temporal coherence of the source. As a result, four distinct interferences appear simultaneously on the multiplexed hologram defined by Eq. (1). The off-axis fringe orientations, as dictated by the angle of the four reference-arm mirrors, can be aligned so that in the spatial-frequency domain of the hologram, each of the cross-correlation elements will be located off-axis and without overlap to any other element. Thus, it is possible to fully reconstruct the four complex wavefronts of the four sample layers without loss of resolution or magnification.

In the case of biological samples, whereby there is no distinct layer structure, the fringe orientation will have to be adjusted manually until exclusive interference orientations are obtained. However, for parallel sample structure, as relevant in metrology, the off-axis angle can be predetermined. The orientation of the off-axis fringes dictated by each reference mirror, positioned in a known axial location, can be used to detect the axial location of each layer. For example, in our case, reference mirror M1 always interferes with layer 1 with 45° fringe orientation, thus the matching frequencies for 45° fringe orientation will always denote the information found in layer 1.

In order to illustrate the discrete fringe orientation from each reference arm, we first placed a single mirror in the sample arm and scanned it axially using an electronic translation stage (T-Cube DC Servo Motor Controller, TDC001, Thorlabs) with axial steps of 10 μm. When the axial location of each of the four reference mirrors matched the axial location of the sample mirror, within the coherence length of the source, an off-axis hologram with fringe orientation corresponding to the off-axis angle of the reference mirror was created. Thus, interference is obtained only when there is path matching from each of the respective mirrors. The first interference occurs as a result of reference mirror M1, and continues through to interference from reference mirror M4. Figure 2 presents a graph of the hologram fringe visibility values, whilst scanning the axial location of the sample mirror. Each distinct peak is obtained from interference with a different mirror, as marked in Figure 2. The visibility of each hologram was determined as the ratio of the first-order spectral peak to that of the zero-order spectral peak. The sample axial location obtained from the electronic scanning stage is shown across the horizontal axis of this graph. As can be seen, four distinct peaks are detected, indicating the axial locations of each layer. Layer one is found at $z = 50$ μm, layer 2 at $z = 125$ μm, layer 3 at $z = 245$ μm, and layer 4 at $z = 315$ μm. The average FWHM of the four visibility curves was calculated to be 24.5 μm.

Employing the axial distances of the maximum fringe visibility indicated by the peaks of Fig. 2, the left images of Fig. 3 show the off-axis holograms obtained at those distances. As can be seen, off-axis interference with distinct fringe orientation is seen in each of these off-axis holograms without interferences from the other $z$ planes. This is also evident in the corresponding power spectra obtained by a 2-D Fourier transform of the hologram, as shown on the right side of Fig. 3, whereby only one pair of cross-correlation elements is seen in each power spectrum. Visualization 1 presents the dynamic off-axis hologram, the corresponding power spectra and the hologram fringe visibility with the instantaneous $z$ location indicated by a red dot, as obtained during scanning the sample mirror. Thus, it was possible to obtain optical sectioning due to the orientation of the fringes at axial locations of maximal visibility.

To allow for simultaneous acquisition of all four layers in a single exposure, we next imaged the four-layered sample described in Fig. 1(b). For a multilayered sample, the position of each reference mirror can be adjusted by the micrometers upon which it sits, for the maximal interference visibility, and we can use the relative distance between each reference mirror to calculate the distance between the layers. The peak distance is determined relative to layer 1, which is located at distance 0.

In this case, the camera recorded the multiplexed hologram shown in Fig. 4(a), which simultaneously contains four off-axis interference fringe orientations. The corresponding power spectrum obtained by a 2-D Fourier transform of the multiplexed hologram is shown in Fig. 4(b). As shown, the unique fringe orientations from the layers create four fully-separated cross correlation pairs. Note, that the power spectrum does not contain cross-correlation elements between nonmatching sample and reference beams as they are located at distances greater than the coherence length of the source. This is mathematically shown in Eq. (1). By employing off-axis hologram reconstruction, whereby the four cross-correlation elements (one from each pair of cross-correlation elements, each located on either side of the spatial spectrum) are digitally cropped and undergo an inverse Fourier transform, we can obtain the complex wavefront, containing both the amplitude and the quantitative phase profile of that corresponding layer. Figure 4(c) shows the reconstructed amplitude (left) and unwrapped phase (right) profiles of each layer. Due to the thickness of the sample (0.28 mm), it is not possible to use the phase profile to detect the axial distance of that layer, as the 2-D unwrapping algorithm would yield similar phase values for each layer. However, the relative distances of the reference mirrors, known in advance, allow for the calculation of the distances between the layers, and the fringe orientation refers to a particular reference mirror. Using layer 1 as our reference zero distance, the remaining axial locations can be calculated. As can be seen in Fig. 4(a), the 'I' had $45^0$ oriented fringes, meaning that it is attributed to the first layer, and located at $z = 0$ μm. The 'N' had $0^0$ (horizontally oriented) fringes, meaning that it is attributed to the second layer, and located at $z = 95.25$ μm. The 'M' had $90^0$ (vertically oriented) fringes, meaning that it is attributed to the third layer, and located at $z = 187.33$ μm. Finally, the 'O' had $135^0$ oriented fringes, meaning that it is attributed to the fourth layer, and located at $z = 273.05$ μm. The distances between the layers correspond with the expected thickness of the coverslips.

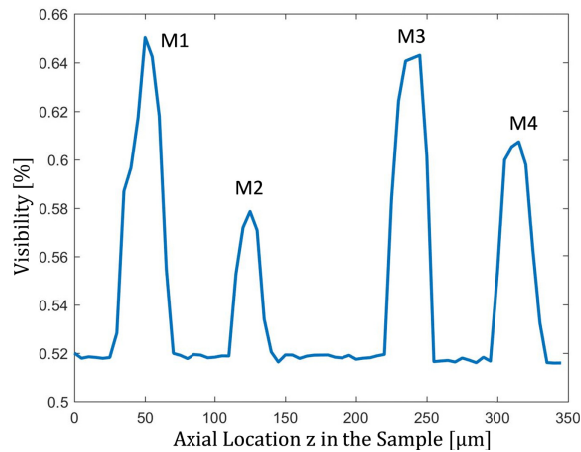

Fig. 2. Hologram fringe visibility as a function of sample thickness due to axial scanning of a sample containing only a single layered mirror. The four distinct peaks indicate the location of each layer in the thick sample, as obtained from interferences with mirrors M1 through M4. Averaged FWHM width of the peaks is 24.5 μm.

To conclude, we presented multiplexing of multiple sections in a low-coherence interferometric setup, which enables the simultaneous amplitude and phase reconstruction of multiple sections in optically thick samples in a single camera acquisition without out-of-focus objects in each section. Whilst we tested the model on a stationary target of four layers, it could be applied to dynamic targets as well. The method is applicable to cases in which the depth of field is large (several hundreds of micrometers, dictating spatial resolution in the order of several tens of micrometer), and thus the entire 3-D sample is in focus on the camera. In these cases, we cannot know the order of layers in the 3-D sample. To solve this problem, we practically divided the depth of field into several layers, each creating a different fringe orientation on the camera simultaneously, and the different layers do not interact with each other due to coherence gating. We expect this method to be useful for optical metrology and biomedical imaging.

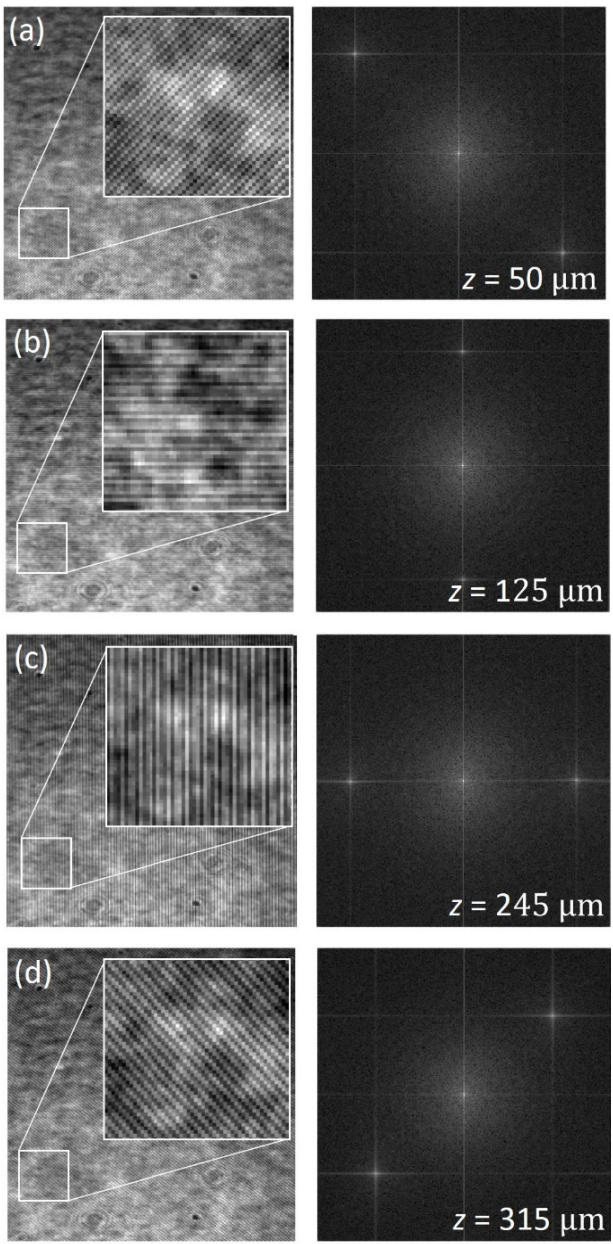

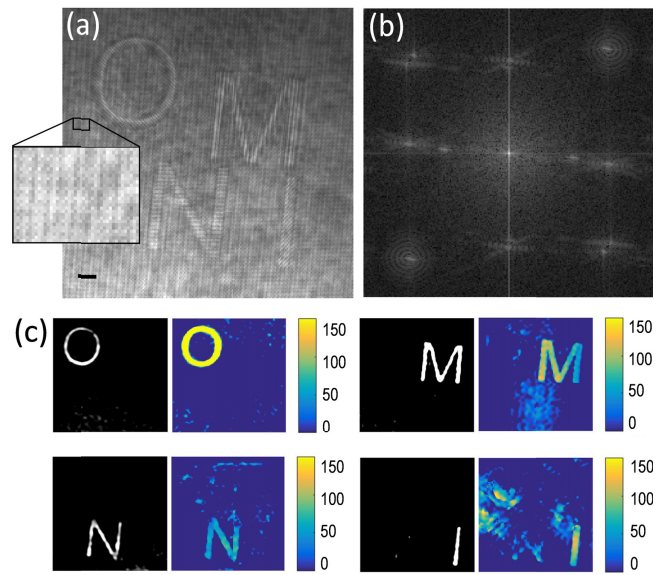

Fig. 4. Optical sectioning obtained by off-axis holographic multiplexing for the four-layer sample described in Fig. 1(b). (a) Multiplexed off-axis hologram acquired in a single camera exposure. Scalebar represents 130 µm on the sample. (b) The corresponding power spectra containing four distinct cross correlation pairs. (c) Reconstructed amplitude (left) and unwrapped phase (right) profiles obtained by cropping the corresponding cross-correlation elements in (b), enabling simultaneous reconstruction of all four layer complex wavefronts from a single camera exposure. Colorbar represents height values in nm.

Fig. 3. The off-axis holograms (left) and their corresponding power spectra (right) obtained at the maximum fringe visibility planes during axial scanning of the sample containing only a single layered mirror. The distinct cross correlation elements are clearly seen in each image. (a) $z = 50$ µm. (b) $z = 125$ µm. (c) $z = 245$ µm. (d) $z = 315$ µm. Visualization 1 presents the continuous axial scanning.

**Funding.** European Research Council (ERC) (678316).